# Real-Time Discrete Fractional Fourier Transform Using Metamaterial Coupled Lines Network

Rasool Keshavarz[1], Negin Shariati[1], Mohammad-Ali Miri[2,3]

*Abstract*— **Discrete Fractional Fourier Transforms (DFrFT) are universal mathematical tools in signal processing, communications and microwave sensing. Despite the excessive applications of DFrFT, implementation of corresponding fractional orders in the baseband signal often leads to bulky, power-hungry, and high-latency systems. In this paper, we present a passive metamaterial coupled lines network (MCLN) that performs the analog DFrFT in real-time at microwave frequencies. The proposed MCLN consists of $M$ parallel microstrip transmission lines (TLs) in which adjacent TLs are loaded with interdigital capacitors to enhance the coupling level. We show that with proper design of the coupling coefficients between adjacent channels, the MCLN can perform an M-point DFrFT of an arbitrary fractional order that can be designed through the length of the network. In the context of real-time signal processing for realization of DFrFT, we design, model, simulate and implement a 16×16 MCLN and experimentally demonstrate the performance of the proposed structure. The proposed innovative approach is versatile and is capable to be used in various applications where DFrFT is an essential tool. The proposed design scheme based on MCLN is scalable across the frequency spectrum and can be applied to millimeter and submillimeter wave systems.**

## I. INTRODUCTION

THE increasing demand for high data rate communication systems needs reliable and fast signal processing units [1]. Digital Signal Processing (DSP) is a foremost solution for modulation and demodulation of signal in wireless communication systems. Regarding the compact size and flexibility, DSP devices are good candidate for complicated signal processing techniques [2]. However, DSP units suffer from poor performance at high frequencies, low speed, high power consumption, quantization error, low spurious free dynamic range (SFDR) at high frequencies, and high cost [3], [4].

Real-time signal processing of RF/microwave signals leverages improvement of temporal and spectral efficiencies to realize fast and reliable radar and communication systems [5]. Therefore, Analog Signal Processing (ASP) as a real-time method is evitable in ultra-fast and highly sensitive processing units [6]. This method denotes a new paradigm for RF/microwave signal processing without requiring signal samplers and digital processors in conventional DSP systems [7]. For instance, real-time Fourier transform (FT) is a powerful tool for range detection in automotive radars to find obstacles around the vehicle [8]. ASP devices are capable to be scaled up to millimeter-wave frequencies (e.g. 60 GHz band) to achieve wider bandwidths in ultra-high data rate systems without any data buffering [5]. In terms of power consumptions, simplicity in implementation, and cost, ASP devices are more desirable than DSP techniques. ASP devices comprised of RF/microwave circuits in which signal propagates through circuits to manipulate it in the time-domain. Passive circuits are designed based on the dispersion characteristics, thereby controlling the wave propagation through transmission lines (TLs).

Recently, Metamaterials and Metasurfaces have been proposed as a promising solution to solve mathematical equations in various fields. These structures can be designed to manipulate electromagnetic waves in a controlled manner. The unique properties can be used to design devices that perform mathematical operations, such as solving partial differential equations (PDEs), Fourier transforms, matrix operation and integral transforms on electromagnetic signals [9], [10], [11].

The Fractional Fourier Transform (FrFT), often regarded as a generalization of Fourier Transform (FT), is a mathematical tool for various applications dealing with signal processing such as in communication systems, microwave sensors, and radar [12], [13]. [14] A new technique is presented in [14] to implement FrFT using SH0 wave computational metamaterials in the space domain, however the study is limited in theoretical scope. In practical applications, one is often interested in discrete counterparts, i.e., Discrete Fourier Transform (DFT) and Discrete Fractional Fourier Transform (DFrFT) [15], [16]. In the optical frequencies, despite several proposals, the realization of a planar on-chip device that performs DFT remains challenging. On the other hand, recently, it is shown that the DFrFT can be realized with photonic waveguide arrays [17], [18] using nonuniform evanescent coupling to emulate a particular lattice model in quantum physics [19]. More recently, it is demonstrated that such a lattice can be implemented in planar waveguide systems to realize an integrated waveguide lens for beam steering applications [20].

[1]RF and Communication Technologies (RFCT) research laboratory, University of Technology Sydney, Ultimo, NSW 2007, Australia, e-mail: Rasool.Keshavarz@uts.edu.au; Negin.Shariati@uts.edu.au

[2]Department of Physics, Queens College of the City University of New York, Queens, New York 11367 USA, e-mail: mohammadali.miri@qc.cuny.edu

[3]Physics Program, The Graduate Center, City University of New York, New York, New York 10016, USA.

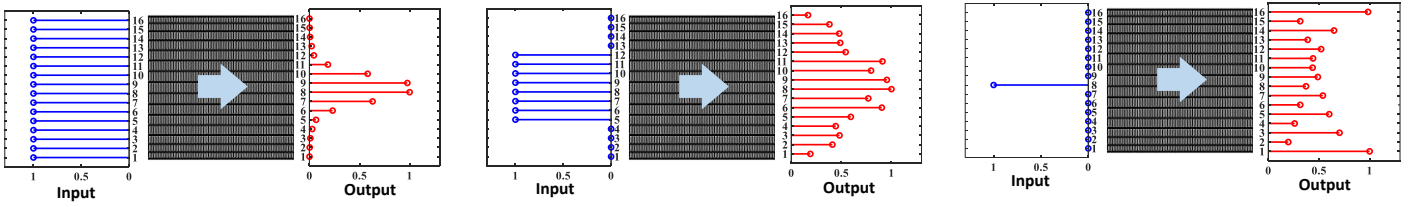

Fig. 1. Schematic illustration of the proposed metamaterial coupled lines network (MCLN) as a real-time Discrete Fractional Fourier Transform (DFrFT) system.

In this paper, a real-time DFrFT system is proposed that works in the microwave frequencies and is scalable to higher and lower frequency ranges. To realize the DFrFT system, a metamaterial coupled lines network (MCLN) is proposed. MCLN consists of $M$ parallel microstrip transmission lines (TLs) that are loaded with interdigital capacitors to improve the coupling coefficients between adjacent TLs. This periodic structure is capable to manipulate the input signal along the propagation direction. Then, the proper signals are delivered to the output ports in such a way that is DFrFT of the input signal (Fig. 1).

Major contributions of this paper are summarized as follows:

- For the first time, the DFrFT is realized in a microwave integrated circuit platform with passive, low-profile and low cost microstrip circuits.

- Interdigital loaded coupled transmission lines have been proposed as a means of achieving a compact and high coupling level MCLN, which can be utilized in the implementation of DFrFT. This technique involves loading interdigital capacitors between coupled transmission lines, resulting in enhanced capacitance and increased coupling between the lines. As a result of interdigital loaded TLs, strong coupling coefficients is achieved between adjacent transmission lines which subsequently results in a smaller and more efficient MCLN, allowing for the implementation of DFrFT in a more compact and practical manner.

- The ratio of length/wavelength in our proposed microwave circuit is by five orders-of-magnitude smaller than the previously reported optical lattice of [17]. This is because of the strong coupling between nearest neighbor lines enabled through the fingering lines which is impossible to implement in the optical frequencies. In [17], for a lattice of 21 waveguides, the length-over-wavelength ratio is estimated to be 74.7mm/632nm ≈ 118196 while in our design, this ratio is 140mm/88mm = 1.59 for a 21-channel lattice in the microwave regime.

- The proposed platform allows for the realization of Jx lattices with much larger number of waveguides (transmission lines) compared to the laser-written glass waveguide discussed in [17]. In fact, the coupling coefficients in the Jx lattice decrease rapidly as moving from the center of the lattice toward the edge channels. As a result, by increasing the number of lattice sites N, the coupling coefficients rapidly approach very small numbers such that the light coupling approaches the noise level. Thus, it is expected that physical realization of the optical lattice discussed in [17] will be limited to only tens of channels. In contrast, in our proposed microwave circuit, one can realize Jx lattices with hundreds of channels while still having significant coupling across all channels. This is again dues to the strong coupling between channels in our system.

- The equivalent circuit model of the proposed MCLN is presented and the achieved good agreement between theory, simulation and measurement results confirms accuracy of the modeling procedure.

- The proposed technique is versatile and can be adopted to different applications where DFrFT is an important tool in signal processing. Further, this technique is scalable to lower and higher frequency ranges.

- Finally, we would like to emphasise on the importance of the proposed realization on an on-chip microwave platform that is relevant to several technological applications. The proposed device is on-chip and can be readily integrated into larger systems for signal processing applications. Apart from opportunities in radio frequency and microwave analog information processing, the proposed circuit can be utilized in microwave quantum integrated circuits for quantum information processing and quantum simulations. This is particularly important considering that superconducting microwave quantum circuits are one of the most important candidates for quantum computing. The proposed design can be readily utilized in the state-of-the-art low-loss superconducting microwave integrated circuit platforms. Furthermore, the design procedure introduced here can be readily applied to millimeter wave circuits that paves the way for other potential applications.

The organization of this paper is as follows: The theory of Jx array as a DFrFT system is presented in Section II. The proposed MCLN is designed, modeled and simulated in Section III. Further, we show that MCLN is a good candidate to realize real-time DFrFT in the microwave frequencies. Section IV includes fabrication, measurement and discussions. Lastly, the conclusion is presented in Section V.

## II. DFrFT AND ITS PHYSICAL IMPLEMENTATION

The Discrete Fractional Fourier Transform (DFrFT) is a generalization of the Discrete Fourier Transform (DFT) to fractional powers. Unlike its continuous counterpart, the DFrFT has different definitions. To introduce the DFrFT, here, we take a step back to look at the DFT and then we discuss its generalization to Fractional Fourier Transform (FrFT) [15]. Considering an N-point complex valued sequence as a vector $\mathbf{x} = (x_0, x_1, x_2, \cdots, x_{N-1})^t)$, its discrete Fourier transform vector is a sequence of the same length $\mathbf{X} = (X_0, X_1, X_2, \cdots, X_{N-1})^t$ defined through $X_k = \frac{1}{\sqrt{N}} \sum_{n=0}^{N-1} x_n e^{-j2\pi kn/N}$, while the original sequence can be reconstructed through the inverse discrete Fourier transform relation $x_n = \frac{1}{\sqrt{N}} \sum_{k=0}^{N-1} X_k e^{j2\pi kn/N}$.

The DFT is a linear unitary operation that can be described through the so-called DFT matrix $F$ according to:

$$\mathbf{X} = F\mathbf{x} \tag{1}$$

where, $F_{k,n} = e^{-j2\pi kn/N}/\sqrt{N}$ ($k, n = 0,1,2,\cdots, N-1$). It is straightforward to show that the DFT matrix $F$ satisfies the following relations, $F^2 = P$ and $F^4 = I$, where $P$ is the parity operator imposing a mirror symmetry and $I$ is the identity matrix. Furthermore, for $N \geq 4$, the DFT matrix has only four distinct eigenvalues $\lambda_m$ ($m = 1, \cdots, N$) of $\lambda_m \in \{1, j, -1, -j\}$ [21], [22], [23]. Therefore, considering a set of orthonormal eigenvector basis $\mathbf{v}_p$ for the DFT matrix $F$, it can be written in an eigen decomposition from as:

$$F = \sum_{n=1}^{N} \mathbf{v}_n e^{-j\frac{\pi}{2}n} \mathbf{v}_n^\dagger \tag{2}$$

(where, "†" is a transpose conjugate).

The DFT is useful in many applications, including signal spectral analysis and understanding how a signal can be expressed as a combination of waves, which allow for manipulation of that signal and comparisons of different signals. Moreover, RF/microwave waves can be filtered using DFT technique to avoid noise into the important components of a signal.

DFrFT has attracted a considerable amount of attention, resulting in many applications in the areas of optics and signal processing. Assuming $\mathbf{v}_n$ to be an arbitrary orthonormal eigenvector set of the $N \times N$ DFT matrix and $e^{-j\frac{\pi}{2}n}$ to be the associated eigenvalues, the DFrFT matrix, $K_\alpha$, can now be defined as:

$$K_\alpha = \sum_{n=1}^{N} \mathbf{v}_n \left(e^{-j\frac{\pi}{2}n}\right)^\alpha \mathbf{v}_n^\dagger \tag{3}$$

where, $\mathbf{v}_n$ are orthonormal set in $\mathbb{R}^N$ and $\alpha$ is the order of DFrFT. As it will be discussed later, this form can be effectively emulated with a lattice model that can be realized with coupled line networks.

To effectively implement the microstrip Jx Array in microwave frequencies, high coupling level coupled line couplers are required. The conventional edged coupled microstrip coupled lines (as shown in Fig. 2) are not practical due to their low coupling level between adjacent transmission lines, leading to a long and inefficient length of the Jx array. To overcome this issue, we propose using interdigital loaded transmission lines as a Metamaterial Coupled Line Network (MCLN).

MCLN is an artificially engineered structure composed of a periodic structure with a coupling network that can manipulate electromagnetic waves in unique and specific ways. The coupling network provides a high coupling level and compact size, while the periodicity of the structure allows for controlled and predictable interaction with electromagnetic waves. This makes MCLN useful in various applications, including optics and microwave engineering. The proposed periodic structure in MCLN can exhibit the desired electromagnetic responses by carefully engineering the coupling coefficients.

In this paper, we propose a design guide for a passive microstrip network that is capable to calculate DFrFT of the input signal in the microwave frequencies, and exhibit it in the output ports. The proposed real-time DFrFT system includes a $M \times M$ metamaterial coupled-lines network (MCLN) as shown in Fig. 2, where $M$ is the number of input/output ports. The proposed MCLN is composed of an array of evanescently-coupled microstrip transmission lines that are loaded with interdigital capacitors to increase the coupling level between adjacent TLs [24], [25]. The proposed MCLN consists of $N$ unit cells which are cascaded to realize a periodic metamaterial structure. Each unit cell should satisfy the homogeneity condition, $p \ll \lambda_g$, where $p$ and $\lambda_g$ are the length of unit cell and guided wavelength of the structure, respectively [26], [27].

Assuming that all TLs in Fig. 2 are impedance-matched to avoid reflection, one can describe the evolution of electromagnetic waves in the system through the coupled-mode formalism as commonly used in photonic waveguide arrays [20]. Hence, in this framework, the forward-propagating normalized voltage in the $i$'th line can be considered as $V_i^+(z) = a_i(z)e^{-j\beta z}$, where $\beta_0$ is the propagation constant of the individual transmission line and $a_i(z)$ is a slowly varying amplitude, i.e., $|da_i/dz| \ll \beta_0$. In this manner, the evolution of the complex wave amplitudes along the propagation direction $z$ can be described through the coupled mode equations [28], [29], [30], [31]:

$$\frac{da_i}{dz} = -j\beta_0 a_i - j\left(\kappa_{i,i+1} a_{i+1} + \kappa_{i,i-1} a_{i-1}\right), \quad j = 1,2,\ldots,M \tag{4}$$

where, $\kappa_{i,i\pm1}$ and $\beta_0$ are the coupling coefficient between two adjacent TLs and propagation constant of isolated lines, respectively. This relation can be cast in matrix form resulting in:

$$\frac{d\mathbf{a}}{dz} = -jH\mathbf{a} \tag{5}$$

where, $H$ is a tri-diagonal coupled mode matrix

$$H = \begin{bmatrix} \beta_0 & \kappa_{1,2} & \cdots & 0 & 0 \\ \kappa_{1,2} & \beta_0 & & 0 & 0 \\ \vdots & & \ddots & & \vdots \\ 0 & 0 & & \beta_0 & \kappa_{N-1,N} \\ 0 & 0 & \cdots & \kappa_{N-1,N} & \beta_0 \end{bmatrix} \tag{6}$$

For a given length, $L_N$, the MCLN is an $M$-port network that can be described through the transfer matrix, $T$, between its inputs ($X$) and outputs ($Y$) as:

$$Y = TX \qquad (7)$$

where, $X = [V_1^+(z=0), \cdots, V_N^+(z=0)]^T$, $Y = [V_1^+(z=L_N), \cdots, V_N^+(z=L_N)]^T$, and the transmission matrix $T$ is given by:

$$T = \exp(-jHL_N) \qquad (8)$$

w For the Jx lattice, the coupling coefficients are given by [20]:

$$\kappa_{i,i+1} = \frac{\kappa_0}{2}\sqrt{(N-i)i} \qquad i = 1,2,\dots,N-1 \qquad (9)$$

where, $\kappa_0$ is a free scaling parameter.

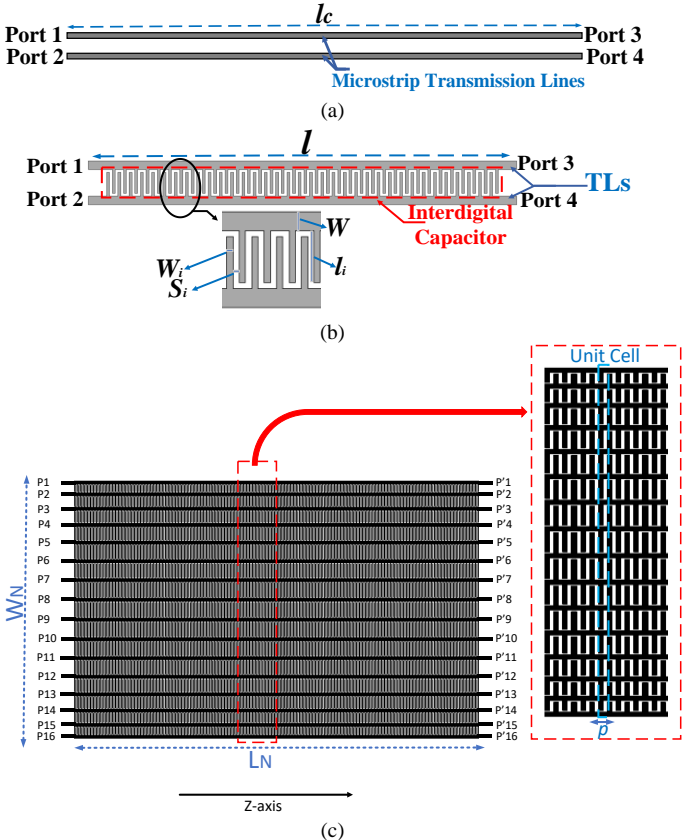

Fig. 2. a) Conventional edged-coupled transmission lines, b) interdigital loaded coupled lines, c) proposed 16×16 Metamaterial Coupled Lines Network (MCLN) with $N$ cells.

The transmission matrix of the proposed MCLN for a given length $L_N$ can be written as:

$$T_{MCLN} = \sum_n e^{-j\lambda_n L_N}\phi_n\phi_n^\dagger \qquad (10)$$

where, $\lambda_n$ is the eigenvalue of the coupled mode matrix $H$ and $\phi_n$ represents the corresponding eigenvectors. The interesting property of the Jx operator is that the eigenvalues are equally spaced as:

$$\lambda_m = \beta_0 + n\kappa_0 \ ; \ \ n = -\frac{N}{2}, \dots, \frac{N}{2} \qquad (11)$$

Therefore, relation (10) can be written as:

$$T_{MCLN} = e^{-j\beta_0 L_N}\sum_n \left(e^{-j\frac{\pi}{2}\alpha}\right)^n \phi_n\phi_n^\dagger \qquad (12)$$

By comparing (3) and (12), $T_{MCLN}$ is clearly a DFrFT with the fractional order:

$$\alpha = \frac{k_0 L_N}{\pi/2} \qquad (13)$$

Therefore, DFrFT of any order can be obtained by the coupling scaling parameter $\kappa_0$ or the array length . It is worth mentioning that the choice of $\alpha = 1$ ($k_0 L_N = \pi/2$) does not result in the DFT. Instead, $\alpha = 1$ in the limit of $N \to \infty$ result in the FT. Therefore, for the choice of $\alpha = 1$, the DFrFT as defined above, can be considered as an approximate DFT. The design procedure of proposed MCLN is summarized in Fig. 3. According to this figure, the first step is finding the maximum coupling level ($k_{max}$) of interdigital loaded coupled lines. Then, after calculating $k_0$ the length of the MCLN is derived ($L_N$).

Fig. 4 exhibits S-parameters (amplitude and unwrap phase) of a $16 \times 16$ MCLN (Fig. 2) for different excitation port number ($i$). Since the proposed structure is symmetry around center line, the S-parameters can be simplified as:

$$for \ j > M/2 \to S_{i,j} = S_{i,(M-j+1)} \qquad i,j = 1,\dots,M \qquad (14)$$

As more input/output ports are added, the overall length of the CLN increases. To achieve a compact structure, a high coupling level is needed between the transmission lines. Conventional coupled lines have low coupling, making the CLN long and difficult to implement. To overcome this, interdigital loaded coupled lines which have high coupling, can be used to create a compact CLN. The interdigital capacitance between adjacent transmission lines boosts mutual coupling ($k_{max}$) resulting in a smaller CLN structure. Moreover, according to (13), for a specific $\alpha$ coefficient, $k_0 L_N$ is constant and by increasing the coupling level ($k_0$) the length of array ($L_N$) decreases.

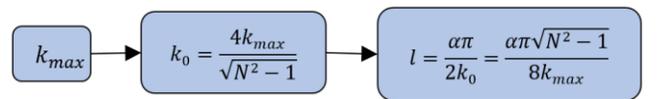

Fig. 3. Design procedure of the proposed MCLN.

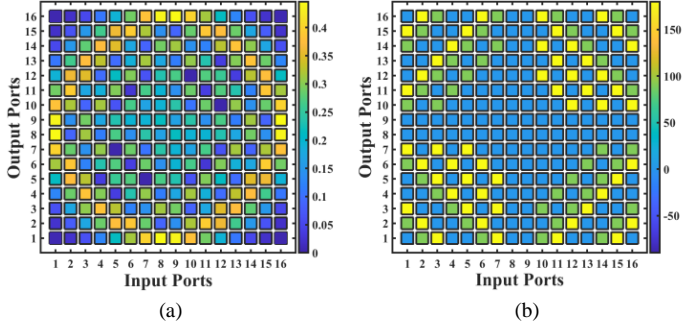

Fig. 4. S-Parameters (amplitude and phase) of the proposed 16×16 MCLN for different input and output ports.

In Section III, we propose a novel technique to design a high coupling level metamaterial coupled line network (MCLN) to achieve a compact array for the proposed DFrFT system.

## III. Design Procedure of MCLN and Simulation Results

The equivalent circuit model of the proposed MCLN is presented in Fig. 5. According to this figure, $L$ and $C$ are inductance and capacitance (per unit length) of each cell, respectively. $C_i$ $(i=1,2,..., M-1)$ is per unit length interdigital capacitor between adjacent TLs.

As shown in Fig. 2, the transmission lines (TLs) are positioned as far apart as possible to minimize any crosstalk between them. The dominant coupling mechanism between adjacent TLs is modeled as interdigital capacitors. The design process starts with the calculation of these capacitances, which is based on equations found in relevant literature [32], [33]. Full-wave simulation and optimization in ADS software are then used to determine the exact values of the interdigital structures, considering a potential for crosstalk between the TLs.

As we discussed in Section II and according to (11), by increasing the coupling level between TLs, the total size of MCLN $(L_N)$ reduces. The interdigital capacitance between adjacent transmission lines boosts mutual coupling $(k_{max})$ resulting in a smaller MCLN structure. However, there is a trade-off between the length $(L_N)$ and width $(W_N)$ of the CLN, for the FR-4 substrate with $\varepsilon_r = 4.3$, thickness of 1.6 mm and $tan(\delta) = 0.03$ as shown in Fig. 6. This figure shows simulation results of a 2-port MCLN using ADS software. The results illustrate that the overall length of coupled lines to achieve maximum forward coupling level at 2.5 GHz is $L_N = 403\ mm\ (3.3\lambda)$ which is not feasible for CLN.

Moreover, to simplify the design procedure, the equivalent circuit model of TLs is assumed lossless. Therefore, the proposed model can be used as an initial step in the design approach. Then, the designed MCLN is optimized using full-wave electromagnetic simulators (ADS) to consider all practical conditions in the simulation procedure.

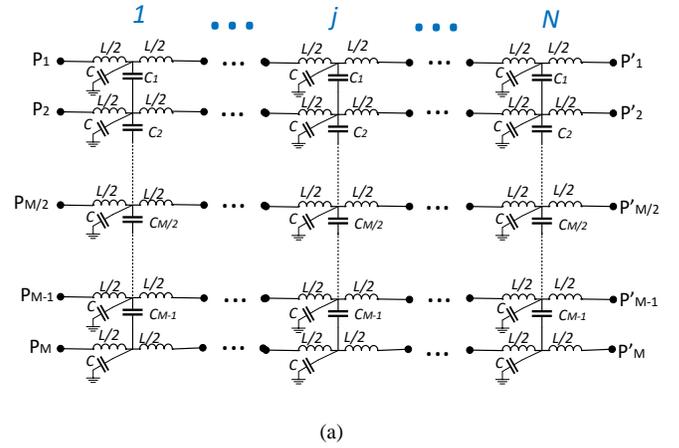

(a)

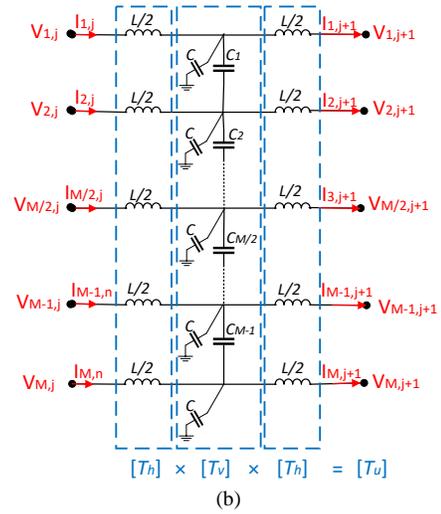

(b)

Fig. 5. Proposed M×M metamaterial coupled lines network (MCLN) with N cells, a) equivalent circuit model of MCLN, b) column unit cells.

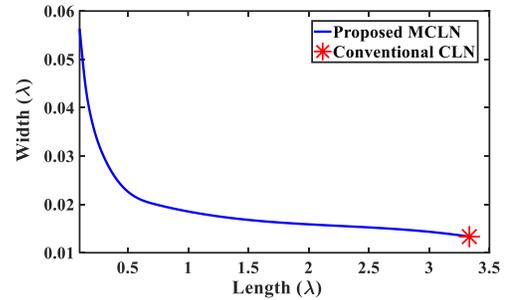

Fig. 6. Trade-off between length $(L_N)$ and width $(W_N)$ of two coupled lines for $N=2$ in $\lambda$.

Fig. 5 shows that the equivalent circuit model of Coupled-Line Network (CLN) is constituted of $N$ column unit cells that are connected to each other as a cascade configuration. Each column unit cell consists of M input/output ports and can be represented by a 2M×2M transmission matrix of unit cell, $[T_u]$.

$$\begin{bmatrix} [V_{in,u}] \\ [I_{in,u}] \end{bmatrix} = [T_u] \cdot \begin{bmatrix} [V_{out,u}] \\ [I_{out,u}] \end{bmatrix} \tag{15}$$

Where $V_{in,u}$, $V_{out,u}$ $I_{in,u}$, and $V_{out,u}$ are input and output voltages and currents vectors of unit cell, respectively. These four vectors are comprised of voltages and currents of $M$ ports in the $j^{th}$ cell (Fig. 5(b)).

$$[V_{in}] = \begin{bmatrix} V_{1,j} \\ V_{2,j} \\ . \\ . \\ . \\ V_{M-1,j} \\ V_{M,j} \end{bmatrix}, \quad [V_{out}] = \begin{bmatrix} V_{1,j+1} \\ V_{2,j+1} \\ . \\ . \\ . \\ V_{M-1,j+1} \\ V_{M,j+1} \end{bmatrix} \quad (16)$$

$$[I_{in}] = \begin{bmatrix} I_{1,j} \\ I_{2,j} \\ . \\ . \\ I_{M-1,j} \\ I_{M,j} \end{bmatrix}, \quad [I_{out}] = \begin{bmatrix} I_{1,j+1} \\ I_{2,j+1} \\ . \\ . \\ I_{M-1,j+1} \\ I_{M,j+1} \end{bmatrix} \quad (17)$$

and

$$[T_u] = \begin{bmatrix} [A] & [B] \\ [C] & [D] \end{bmatrix} = [T_h] \cdot [T_v] \cdot [T_h] \quad (18)$$

Where $[T_h]$ and $[T_v]$ are transmission matrices of horizontal and vertical sections of the equivalent circuit model of unit cell. Further, according to Fig. 5(b), $[T_h]$ and $[T_v]$ are calculated as:

$$[T_h] = \begin{bmatrix} [I] & (j\omega L/2)[I] \\ [O] & [I] \end{bmatrix} \quad (19)$$

$$[T_v] = \begin{bmatrix} [I] & [O] \\ [C_v] & [I] \end{bmatrix} \quad (20)$$

where $[I]$ and $[O]$ represent the $M \times M$ unit and zero matrices, respectively and $[C_v]$ is given by:

$$[C_h] = \begin{bmatrix} Y_{a1} & Y_{b2} & 0 & . & . & . & 0 & 0 & 0 \\ Y_{b2} & Y_{c2} & Y_{b3} & . & . & . & 0 & 0 & 0 \\ 0 & Y_{b3} & Y_{c3} & . & . & . & 0 & 0 & 0 \\ . & . & . & . & & . & . & . \\ . & . & . & & . & . & . \\ . & . & . & & & . & . & . \\ 0 & 0 & 0 & . & . & . & Y_{cM-2} & Y_{bM-1} & 0 \\ 0 & 0 & 0 & . & . & . & Y_{bM-1} & Y_{cM-1} & Y_{bM} \\ 0 & 0 & 0 & . & . & . & 0 & Y_{bM} & Y_{aM} \end{bmatrix} \quad (21)$$

and

$$\begin{cases} Y_{ai} = Y_p + Y_i & i = 1, M \\ Y_{bi} = -Y_i & i = 2 : (M-1) \\ Y_{ci} = Y_p + Y_i + Y_{i-1} & i = 2 : (M-1) \\ Y_p = j\omega C \\ Y_i = j\omega C_i & i = 1 : (M-1) \end{cases} \quad (22)$$

The transmission matrix of the total MCLN constituted of $N$

columns is then simply obtained by taking the $N^{th}$ power of the transmission matrix of the column cell.

$$[T_{MCLN}] = [T_u]^N = \begin{bmatrix} [A_{MCLN}] & [B_{MCLN}] \\ [C_{MCLN}] & [D_{MCLN}] \end{bmatrix} \quad (23)$$

Finally, according to $[T_{MCLN}]$ matrix, we can calculate $[Z_{MCLN}]$ and $S_{MCLN}$ matrices [33]:

$$[Z_{MCLN}] = \begin{bmatrix} [A_{MCLN}] \cdot [C_{MCLN}]^{-1} & [C_{MCLN}]^{-1} \\ [C_{MCLN}]^{-1} & [C_{MCLN}]^{-1} \cdot [D_{MCLN}] \end{bmatrix} \quad (24)$$

$$[S_{MCLN}] = \left[ \frac{[Z_{MCLN}]}{z_0} + I \right] \cdot \left[ \frac{[Z_{MCLN}]}{z_0} - I \right]^{-1} \quad (25)$$

Therefore, S-parameters of the proposed MCLN can be determined based on (23). In the design procedure, the equivalent circuit model components can be calculated using MARLAB software to solve matrix equations. As an example, a 16×16 MCLN is designed at 2.5 GHz based on the proposed equivalent circuit model and design approach. Fig. 7 shows |S|-parameters of output ports against frequency based on the theory. According to this figure and Fig. 4, the output amplitudes validate the performance of the MCLN as a real-time DFrFT system.

Simulation results of the longitudinal evolution of the E-field amplitudes are presented in Fig. 8 for different excitation port. There is a great correlation between theory and simulations, while the E-field patterns in this figure follow the results in Fig. 4.

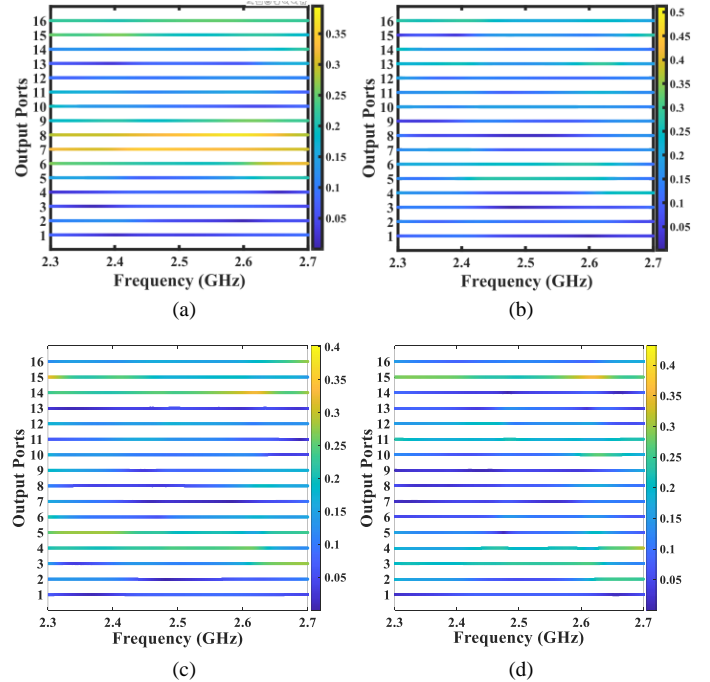

(a)

(b)

(c)

(d)

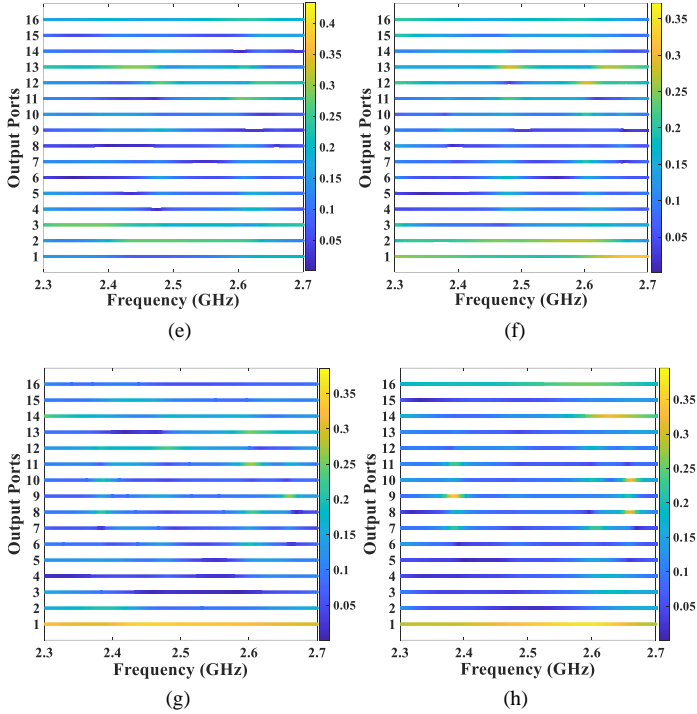

Fig. 7. |S|-parameters of 16×16 MCLN at different output ports based on the design equations (theory) for different excitation port numbers (i), a) *i=1 or 16*, b) *i=2 or 15*, c) *i=3 or 14*, d) *i=4 or 13*, e) *i=5 or 12*, f) *i=6 or 11*, g) *i=7 or 10*, h) *i=8 or 9*.

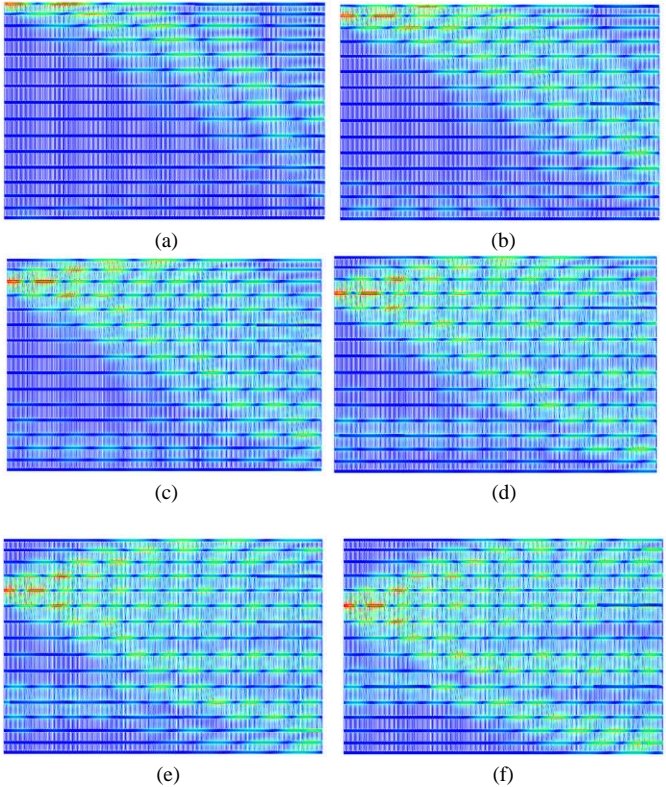

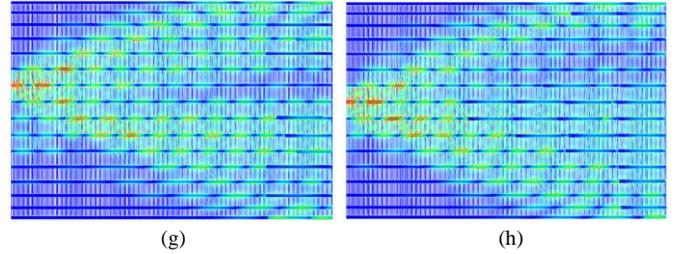

Fig. 8. Simulation results of the longitudinal evolution of the E-field amplitudes through the 16×16 MCLN, for different excitation port numbers (i), a) *i=1 or 16*, b) *i=2 or 15*, c) *i=3 or 14*, d) *i=4 or 13*, e) *i=5 or 12*, f) *i=6 or 11*, g) *i=7 or 10*, h) *i=8 or 9*.

## IV. Fabrication, Measurement and Discussion

To verify the performance of the proposed MCLN as a real-time DFrFT system, a 16×16 MCLN is fabricated on FR-4 substrate with $\varepsilon_r = 4.3$, thickness of 1.6 mm and $tan(\delta) = 0.03$ (Fig. 9). The dimensions of the fabricated MCLN are $L_N$=110 mm and $W_N$=80 mm ($1.25\lambda_g \times 0.9\lambda_g$).

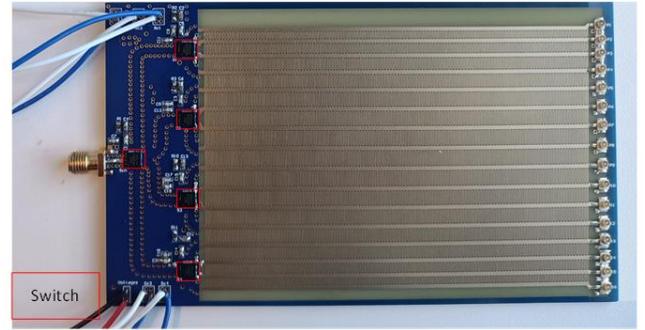

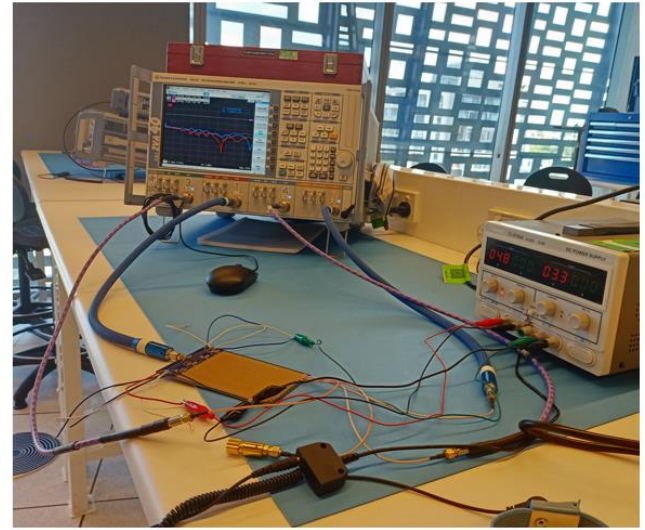

Fig. 9. a) Fabricated prototype on FR-4 substrate with $\varepsilon_r = 4.3$, thickness of 1.6 mm and $tan(\delta) = 0.03$, b) measurement setup.

To measure the |S|-parameters of the fabricated 16×16 MCLN, five absorptive SP4T switches (HSWA4-63DR+) are considered in the input ports to select the proper excitation ports in the input. Moreover, a 4-port vector network analyzer (VNA-ZVA40) is used to measure S-parameters while unconnected output ports are loaded with 50Ω. The measurement setup is shown in Fig. 9 (b).

The measured results of S-parameters for different excitation ports are presented in Fig. 10. According to this figure, the proposed real-time DFrFT circuit operates around 2.5 GHz and the achieved results validate the design procedure of the MCLN.

Fig. 11 displays the measured reflection coefficient of a 16x16 Metamaterial Coupled Line Network (MCLN) at various input ports. Due to the symmetry of the structure, only the reflection coefficients of ports 1 to 8 are shown. The results demonstrate excellent input port matching ($<-10$ dB) across a wide frequency range, which validate the design methodology of the proposed MCLN in terms of matching conditions.

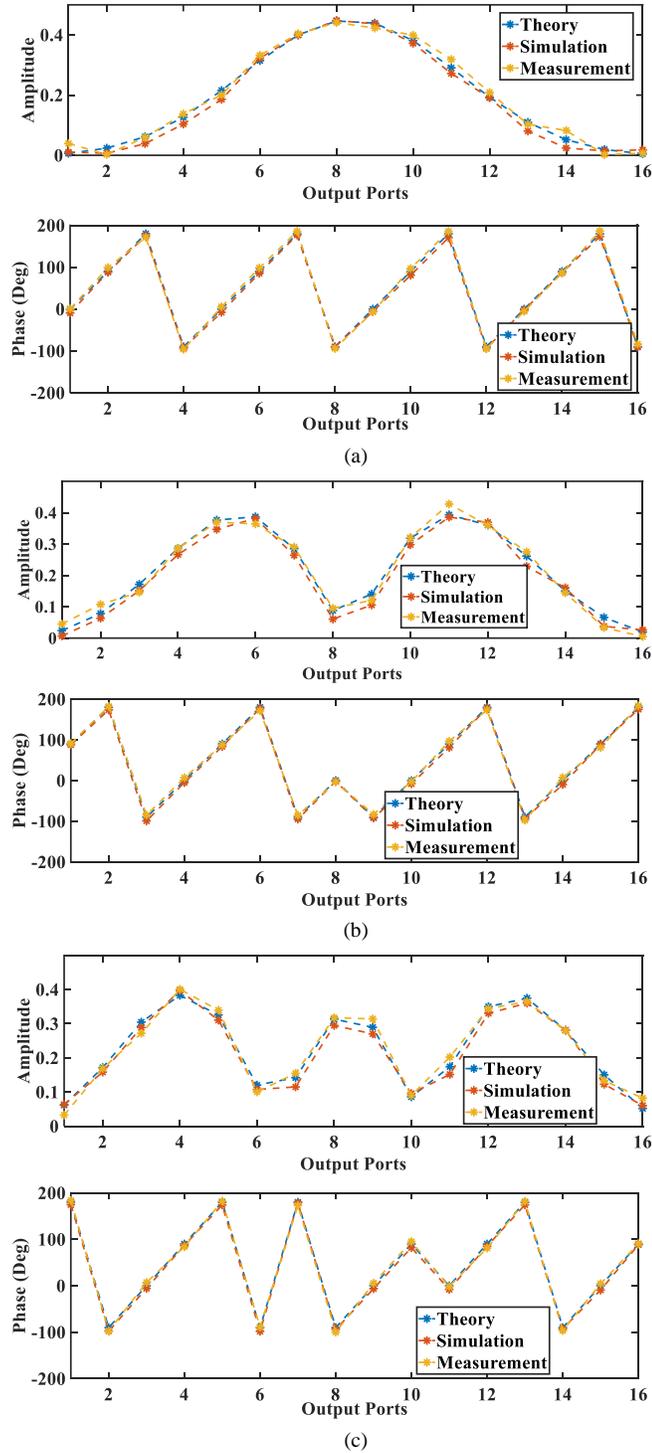

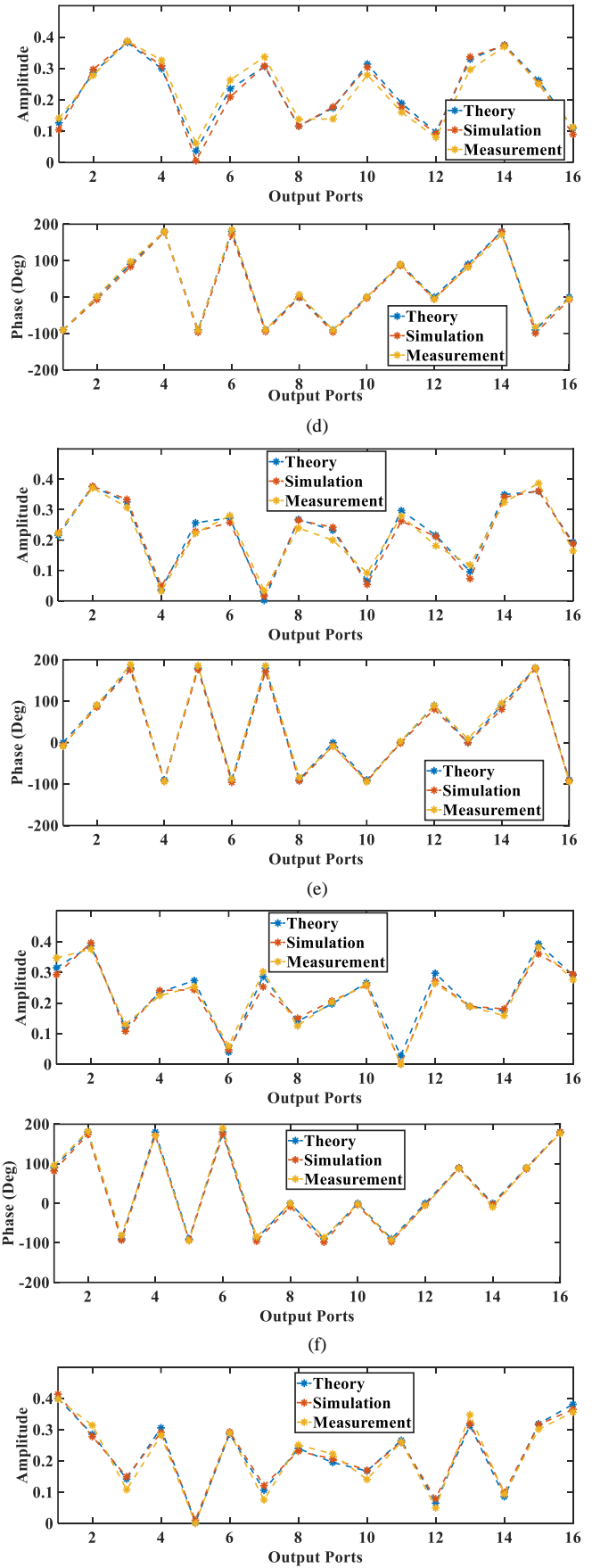

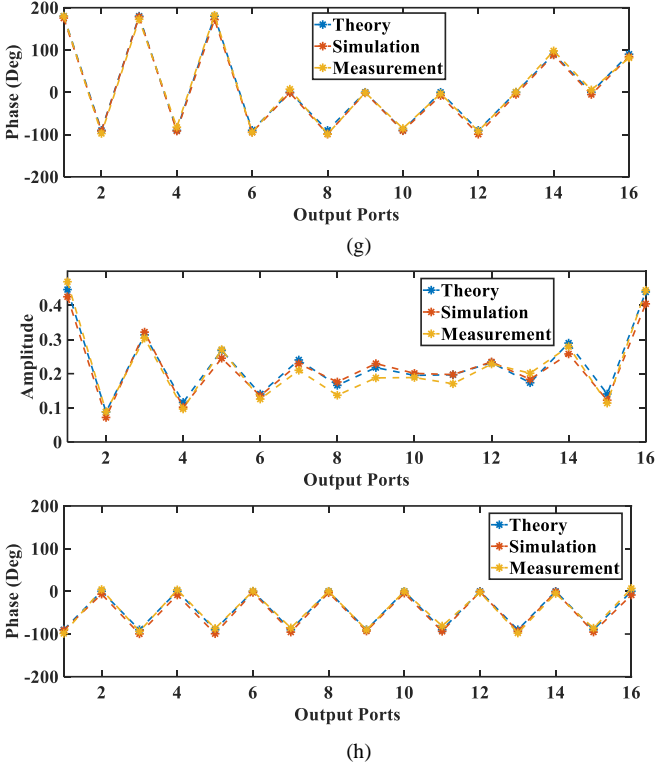

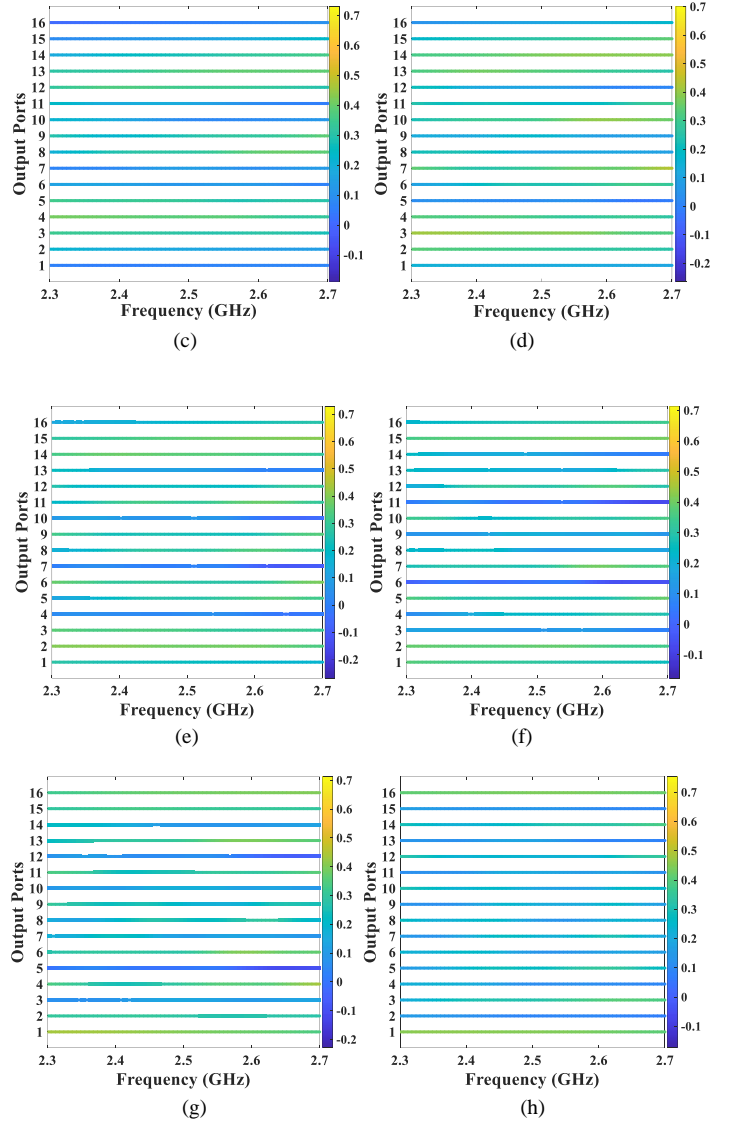

Fig. 10. Theory, simulation and measured S-parameters (amplitude and phase) of 16×16 MCLN for different excitation ports (*i*) at 2.5 GHz, a) *i=1 or 16*, b) *i=2 or 15*, c) *i=3 or 14*, d) *i=4 or 13*, e) *i=5 or 12*, f) *i=6 or 11*, g) *i=7 or 10*, h) *i=8 or 9*.

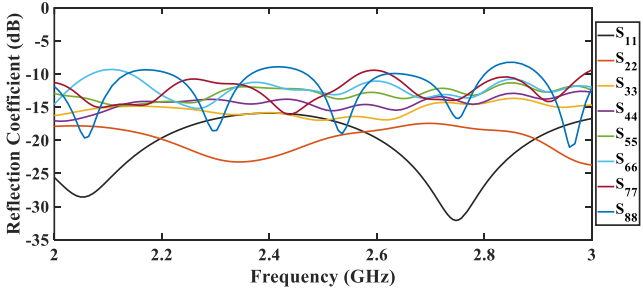

Fig. 11. Measurement results of reflection coefficient at each input port.

Fig. 12 exhibits |S|-parameters of 16×16 MCLN at different output ports for different excitation ports. As can be seen, the measured results follow the desired pattern in Fig. 3.

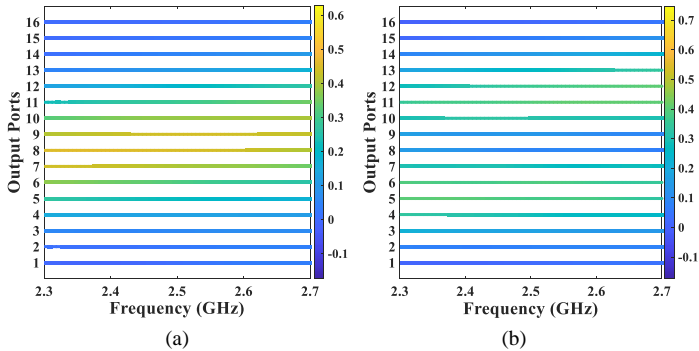

Fig. 12. Measured |S|-parameters of 16×16 MCLN at different output ports for different excitation port number (i) against frequency, a) *i=1 or 16*, b) *i=2 or 15*, c) *i=3 or 14*, d) *i=4 or 13*, e) *i=5 or 12*, f) *i=6 or 11*, g) *i=7 or 10*, h) *i=8 or 9*.

Finally, to validate the performance of the proposed MCLN as a DFrFT circuit, the simulated and measured results for different input signals are presented in Fig. 13. This figure illustrates the amplitude and phase of the DFrFT of two different signals. According to this figure, there is a good agreement between theory, simulation and measurement results, which validates the design process of the proposed DFrFT circuit.

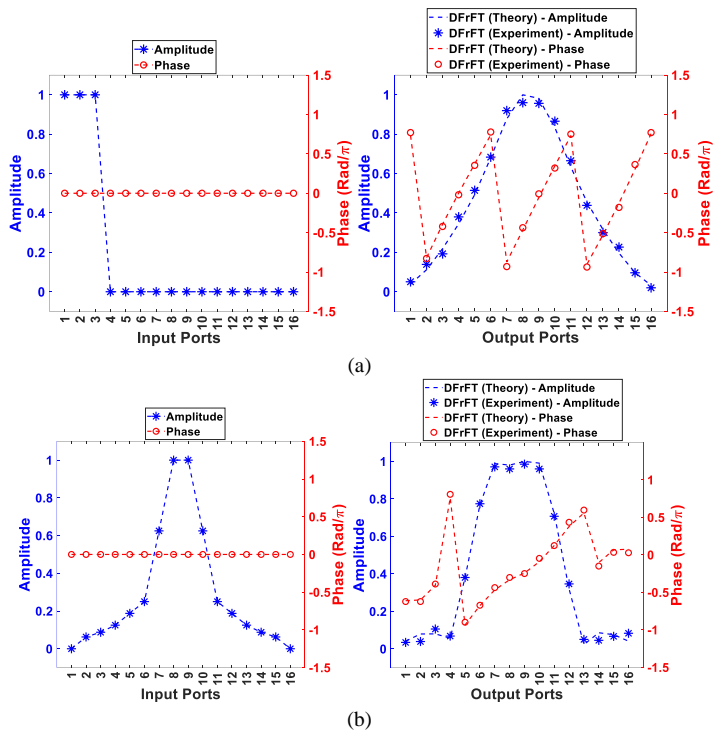

Fig. 13. DFrFT of two different input signals, a) Pulse, b) Gaussian signals.

## V. Conclusion

In this paper, for the first time, a passive metamaterial coupled lines network (MCLN) is designed, fabricated and measured to realize Discrete Fractional Fourier Transform (DFrFT) in microwave frequencies. This method enables a frequency scalable design procedure by parallel microstrip transmission lines (TLs) and loaded interdigital capacitors, which can be applied to millimeter and submillimeter wave systems. To prove the concept, we designed, modeled and fabricated a 16×16 MCLN. The achieved consistency of the measurement and simulation results with theoretical equations based on the equivalent circuit model, demonstrates the effectiveness of the proposed method. This universal design approach is capable to be used in signal processing, communications and microwave sensing. The proposed structure is compact and passive, and hence, is a good candidate for real-time signal processing in microwave frequencies. The proposed approach is versatile and capable to be used in various applications where DFrFT is an important tool in signal processing.

## Acknowledgment

This project is supported by the U.S. Air Force Office of Scientific Research (AFOSR) Young Investigator Program (YIP) Award# FA9550-22-1-0189.

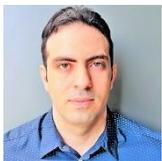

**Rasool Keshavarz** was born in Shiraz, Iran in 1986. He received the Ph.D. degree in Telecommunications Engineering from the Amirkabir University of Technology, Tehran, Iran in 2017 and is currently working as Senior Research Fellow in RFCT Lab at the University of Technology, Sydney, Australia. His main research interests are RF and microwave circuit and system design, sensors, antenna design, wireless power transfer (WPT), and RF energy harvesting (EH).

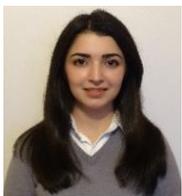

**Negin Shariati** is a Senior Lecturer in the School of Electrical and Data Engineering, Faculty of Engineering and IT, University of Technology Sydney (UTS), Australia. She is also the Director of Women in Engineering at IT (WiEIT) with the Faculty of Engineering and IT. She established the state of the art RF and Communication Technologies (RFCT) research laboratory at UTS in 2018, where she is currently the Co-Director and leads research and development in RF Technologies, Sustainable Sensing, Energy Harvesting, Low-power Internet of Things and AgTech. She leads the Sensing Innovations Constellation at Food Agility CRC (Corporative Research Centre), enabling new innovations in agriculture technologies by focusing on three key interrelated streams; Sensing, Energy and Connectivity.

Since 2018, she has held a joint appointment as a Senior Lecturer at Hokkaido University, externally engaging with research and teaching activities in Japan.

Negin Shariati was the recipient of Standout Research Award in the 2021 IoT Awards Australia, and IEEE Victorian Section Best Research Paper Award 2015. She attracted over four million dollars worth of research funding across a number of CRC and industry projects, where she has taken the lead CI (Chief Investigator) role and also contributed as a member of the CI team.

Dr Shariati completed her PhD in Electrical-Electronics and Communication Technologies at Royal Melbourne Institute of Technology (RMIT), Australia, in 2016. She worked in industry as an Electrical-Electronics Engineer from 2009-2012. Her research interests are in RF and microwave Circuits and Systems, RF Energy Harvesting, Simultaneous Wireless Information and Power Transfer, Sensors and Antennas.

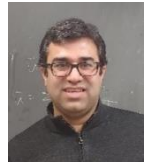

**Mohammad-Ali Miri** is an Assistant Professor of Physics at Queens College and the Graduate Center of the City University of New York. He earned his B.S. and M.S. degrees in Electrical Engineering from Shiraz University (2008), and Sharif University of Technology (2010), and his Ph.D. in Optics from CREOL, the Center for Optics and Photonics, at the University of Central Florida in 2014. Before joining CUNY, he worked as a Postdoctoral Fellow in the Department of Electrical Engineering of the University of Texas at Austin. Dr. Miri's research interests are in the broad areas of optics and photonics, nonlinear optics, optical computing, and integrated photonics. He has authored and co-authored more than 100 publications in peer reviewed journals and conference proceedings, including several highly cited articles on the theory and applications of non-Hermitian photonic systems. Dr. Miri is a Senior Member of Optica, and a recipient of the 2022 Young Investigator Research Program (YIP) award of the Air Force Office of Scientific Research (AFOSR).